# Kinkless electronic junction along one dimensional electronic channel

*Qirong Yao, Jae Whan Park, Choongjae Won, Sang-Wook Cheong, and Han Woong Yeom\**

Dr. Q. Yao, Dr. J. W. Park,
Center for Artificial Low Dimensional Electronic Systems, Institute for Basic Science (IBS), Pohang 37673, Korea

Dr. C. Won
Laboratory for Pohang Emergent Materials, Department of Physics, Pohang University of Science and Technology, Pohang 37673, Korea; Max Plank Pohang University of Science and Technology (POSTECH) Center for Complex Phase Materials, Pohang University of Science and Technology, Pohang 37673, Korea

Prof. S. W. Cheong
Laboratory for Pohang Emergent Materials, Department of Physics, Pohang University of Science and Technology, Pohang 37673, Korea; Max Plank Pohang University of Science and Technology (POSTECH) Center for Complex Phase Materials, Pohang University of Science and Technology, Pohang 37673, Korea; Rutgers Center for Emergent Materials and Department of Physics and Astronomy, Rutgers University, Piscataway, New Jersey, USA

Prof. H. W. Yeom
Center for Artificial Low Dimensional Electronic Systems, Institute for Basic Science (IBS), Pohang 37673, Korea; Department of Physics, Pohang University of Science and Technology, Pohang 37673, Korea
*Email: yeom@postech.ac.kr

Here we report the formation of type-A and type-B electronic junctions without any structural discontinuity along a well-defined 1-nm-wide one-dimensional electronic channel within a van der Waals layer. We employ scanning tunneling microscopy and spectroscopy techniques to investigate the atomic and electronic structure along peculiar domain walls formed on the charge-density-wave phase of 1T-TaS$_2$. We find distinct kinds of abrupt electronic junctions with discontinuities of the band gap along the domain walls, which do not have any structural kinks and defects. Our density-functional calculations reveal a novel mechanism of the electronic junction formation; they are formed by a kinked domain wall in the layer underneath through substantial electronic interlayer coupling. This work demonstrates that the interlayer electronic coupling can be an effective control knob over several-nanometer-scale electronic property of two-dimensional atomic monolayers.

# 1. Introduction

Interlayer interactions in van der Waals (vdW) materials are of prime current interest,[1-3] especially for the consequences of the frustrated interlayer stacking order.[4-9] The most widely investigated example is twisted stacking, which causes a wide range of Moire superstructures and subsequently modulates electronic band structures to result in novel electronic properties. In twisted bilayer graphene at a magic angle, correlated insulator phases were observed at the half-band filling,[4] and superconductivity occurs upon the hole doping.[5] Unconventional electronic states have also been demonstrated in twisted bilayer transition metal dichalcogenides near their low-energy flat bands.[6] In addition, artificial one-dimensional (1D) electronic channels were observed in bilayer tungsten ditelluride with a certain twisted angle.[10]

On the other way around, the layer stacking order can also be perturbed by the presence of grain boundaries or domain walls (DWs) within a vdW layer.[11-13] One particularly interesting case is DWs existing in a Mott-insulating 1T-TaS$_2$ with a commensurate charge density wave (CDW), which has been intensively studied because of their strong electron and spin correlations for quantum physics and applications.[14-18] Both structural configurations and electronic properties of DWs were well-documented recently in this material.[19] Moreover, different methods have been successfully utilized to create DWs by electrical or laser pulses[20,21] and the tuning of the electronic states of DWs themselves by the chemical adsorption[22] were demonstrated. The connection of DWs and the interlayer interaction has also been suggested; the frustrated interlayer coupling in different domains within a DW network was suggested to explain the local coexistence of a metallic and insulating domains.[20]

While most of these works have focused on the properties of a single DW and the domain-to-domain interlayer coupling, in the present work, we open up the discussion on the direct interlayer coupling between DWs. To the best of our knowledge, there has been no explicit discussion of DW-to-DW interlayer coupling in a vdW system while this issue has been investigated in magnetic multilayers.[23] Here, we introduce that the interlayer stacking of DWs and its frustration in 1T-TaS$_2$ can induce novel 1D electronic states. The evidence of the vertical stacking of DWs in neighboring CDW layers of 1T-TaS$_2$ is provided by scanning tunneling microscopy (STM). Spectroscopic measurements clearly identify a few different kinds of abrupt electronic junctions with strong discontinuities of the band gap along the stacked DWs, which do not have any structural kinks and defects within the top layer. Our density-functional calculations reveal that the electronic junctions are formed by a substantial electronic interlayer coupling with a kinked domain wall in the layer underneath, which frustrates the DW-to-DW stacking order. This work suggests that the interlayer electronic coupling can effectively control nanometer-scale electronic properties within two-dimensional atomic monolayers.

## 2. Results and discussion

### 2.1. Surface Morphology of Stacked DWs

**Figure 1**a shows a large-scale STM topographic image of the 1T-TaS$_2$ surface, which is acquired at 4.4 K. Together with regular CDW unitcells appearing as bright clusters, a DW is observed with linearly-arranged clusters of brighter contrast, which gives a translative phase shift of CDW clusters between the left- and right-side domains. According to the classification scheme based on the size and the direction of the phase shift,[14,19] this DW belongs to type DW-2 (see supplementary Figure S1), which is the most popular and most energetically stable one at low temperature. One can further notice that the DW contains three structural kinks within this image, whose atomic and electronic structures were detailed recently.[24] We name straight segments of this DW as DW-2A, DW-2B, DW-2C, and DW-2D, from the top to the bottom, respectively. When a smaller sample bias was applied for scanning at the same region (from +0.4 to +0.2 V for Figure 1a and Figure 1b, respectively), the topographic contrast is subtly altered to reveal more details of the DW. In particular, an extra row of CDW clusters becomes bright on the left side of DW-2A and the lower half of DW-2C as highlighted in Figure 1c and 1d, respectively. Through these changes, the corresponding DW segments apparently become wider. It is worth noting that DW-2B and the upper half of DW-2C remain unchanged at the lower bias, suggesting that the observed contrast changes are not due to the scanning tip.

A closer look into the unitcells with altered contrast, however, could not find any change in the atomic or CDW structure in the top layer. The local contrast change cannot either be related to the kink electronic states, which were characterized in detail to be strongly localized on the kink sites.[24] Thus, we postulate that the observed change may be caused by an electronic effect near the Fermi energy, which is related to objects within the sublayer. We thus scrutinize the corresponding electronic states. In Figure 1e, d$I$/d$V$ curves of the tunneling current, which are proportional to local electronic density of states (LDOS), are displayed for two neighboring domains of the DW. Both of them present the same gap size of about 390 meV. Since the gap size was shown to depend strongly on the interlayer stacking,[9,22] we can confirm that these domains have the same stacking order. The measured gap size corresponds to that of the so-called T$_A$-stacking, where the CDW unitcells of the top and the second layers are stacked center-to-center. This is, however, unexpected since the interlayer stacking of unitcells in one domain must be shifted when one crosses a DW (the left and middle panels in Figure 1f). The consistent interlayer stacking (band gap) across a top-layer DW can only be possible when an identical DW runs immediately below it. Along a straightforward line of thought, if there is a difference between kink positions of DWs in the top and the second layer, one can have a few CDW unitcells nearby the kinks in a different stacking condition (a different interlayer electronic coupling). In particular, for the DW shown in Figure 1, we suspect that the second-layer DW has its kinks shifted from those of the top-layer DW as indicated by the white solid line in order to explain the change of the apparent width of the top-layer DW.

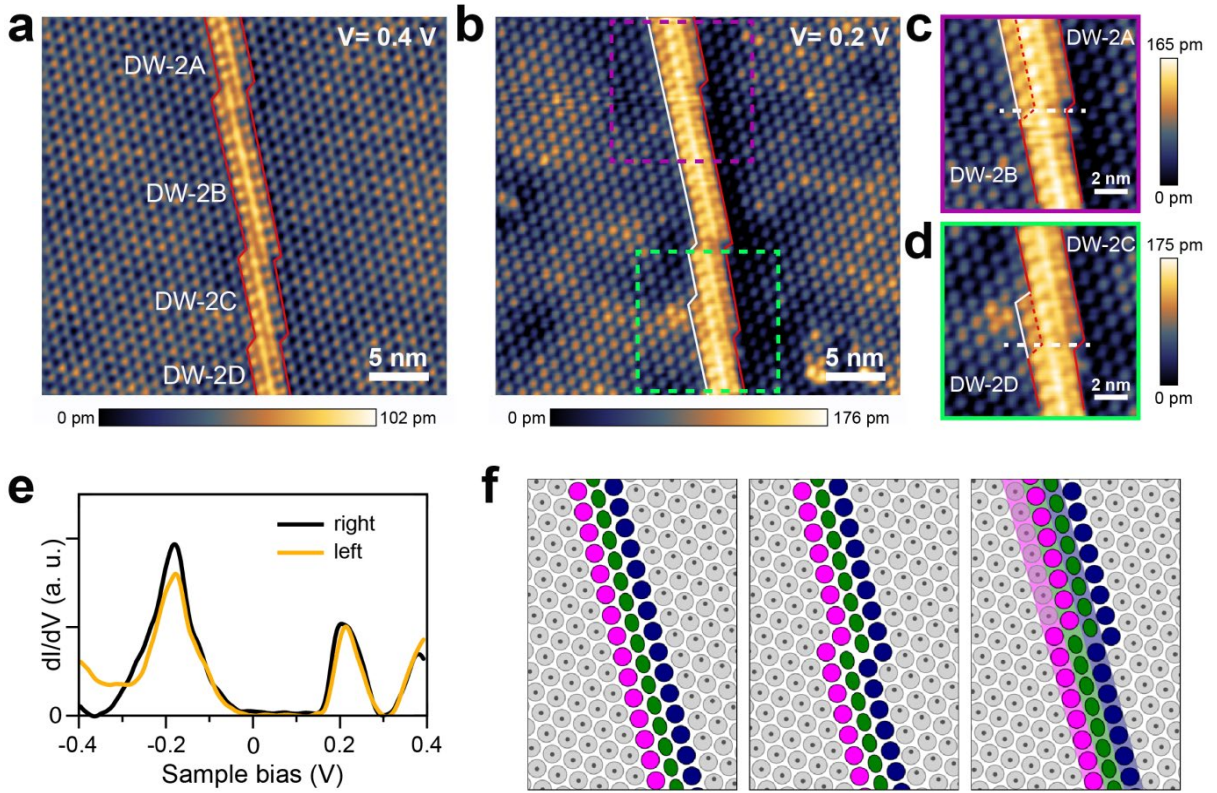

**Figure 1.** STM topography images of domain wall in 1T-TaS$_2$. (a) Sample bias $V_b$ = 0.4 V; (b) $V_b$ = 0.2 V. (c & d) Enlargement of the rectangular section highlighted in (b). (e) Scanning tunneling spectroscopy (STS) spectra of the right and left domains. (f) Schematic configurations of the bilayer DW-2 domain walls. Gray, green, pink, and navy circles denote the CDW clusters in domain, domain wall, left and right domain boundaries, respectively. Dot denotes the central site of the commensurate CDW (CCDW) cluster in the sublayer. The shade lines in the right panel denote the sublayer domain wall and their domain boundaries.

**2.2. Origin of DW-DW Stacking Order**

The above model with two stacked DWs is tested by DFT calculations. Prior to investigating the interaction between DWs in neighboring layers, we briefly overview the atomic and electronic structure of an isolated DW-2 in a single layer. DW-2 is composed of a row of imperfect CDW clusters with a total of eleven Ta atoms each (**Figure 2**a and supplementary Figure S2). DW clusters have topological in-gap states due to their broken CDW unitcell (four Ta atoms in a linear chain and seven in a hexagon). The CDW clusters in the right-side domain boundary (DB, CDW unitcells along the edge of a domain, which neighbor DW clusters) maintain their David-star structure but one of the outer Ta atoms in the cluster is hybridized electronically with the neighboring DW cluster (represented by the connected bonds in Figure 2a). As a result, these DB clusters also exhibit its own in-gap states; its lower Hubbard state shifted upward into the empty state (Supplementary Figure S3) within the gap.[14] In contrast, the CDW clusters in the left-side DB (indicated by a pink dot in Figure 2a) have no such hybridization and maintain their lower and upper Hubbard states except for a small energy shift due to the local band bending caused by the DW. Namely, the single-layer model predicts two rows of clusters (DW and right-side DB) with in-gap electronic states. This is not consistent with the observation of three bright rows in DW-2 in Figure 1b for the bias within the band gap. The model is then elaborated with a double layer with a DW in each layer. Figure 2a depicts the structure of the top layer in the bilayer DW-2, where identical sublayer and DW are located underneath without any translation.

The optimized atomic structure of the DW is similar to that of a single-layer structure, but one Ta atom on the edge of a CDW cluster along the left-side of DB has extra distortion similar to that on the right-side DB (see supplementary Figure S2a and S2b). In the single-layer case, the presence of an odd number of electrons (11 Ta atoms) in the DW leads to strong hybridization with one of DB clusters, resulting in the formation of an energetically stable insulating ground state with an even number of electrons.[19] Conversely, in the bilayer case with a doubling unit cell structure, the DW states symmetrically interact both DBs. Such the interaction leads to an upward band shift of the DB near the Fermi level by charge transfer to the DW to open a correlated gap (see supplementary Figure S3b).[22] The correlated gap is smaller than the Mott gap of the CCDW domain, thus the UHB state of the DB, as an in-gap state, makes the corresponding clusters bright in the STM image simulations (the middle panel in Figure 2b).

While the detailed origin of in-gap states in the right and left DBs differ from the single-layer case (see supplementary Figure S3 and details below), the three bright rows, the DW and both DBs, are in good agreement with the observed STM image of the DW-2B at a sample bias of 0.2 V (the bottom panel in Figure 2b). We then consider the case of the second layer DW shifted by a single CDW unit vector ($a_{CCDW}$, Figure 2c). In this case, the clusters of the top-layer DW couple with the DB clusters of the sublayer DW. An extra row of normal CDW clusters in the left domain (indicated by leftmost blue dots in Figure 2) interacts with the electrons in the bottom layer DW clusters. This electronic coupling makes those clusters develop an in-gap state and become brighter in the simulated empty-state STM image (Figure 2c upper panel). This can thus explain the addition of an extra bright row to DW-2A observed in Figure 1. It turns out that the simulated STM images are in good agreement with the experimental results (Figure 2c), which supports our model of the DW stacking.

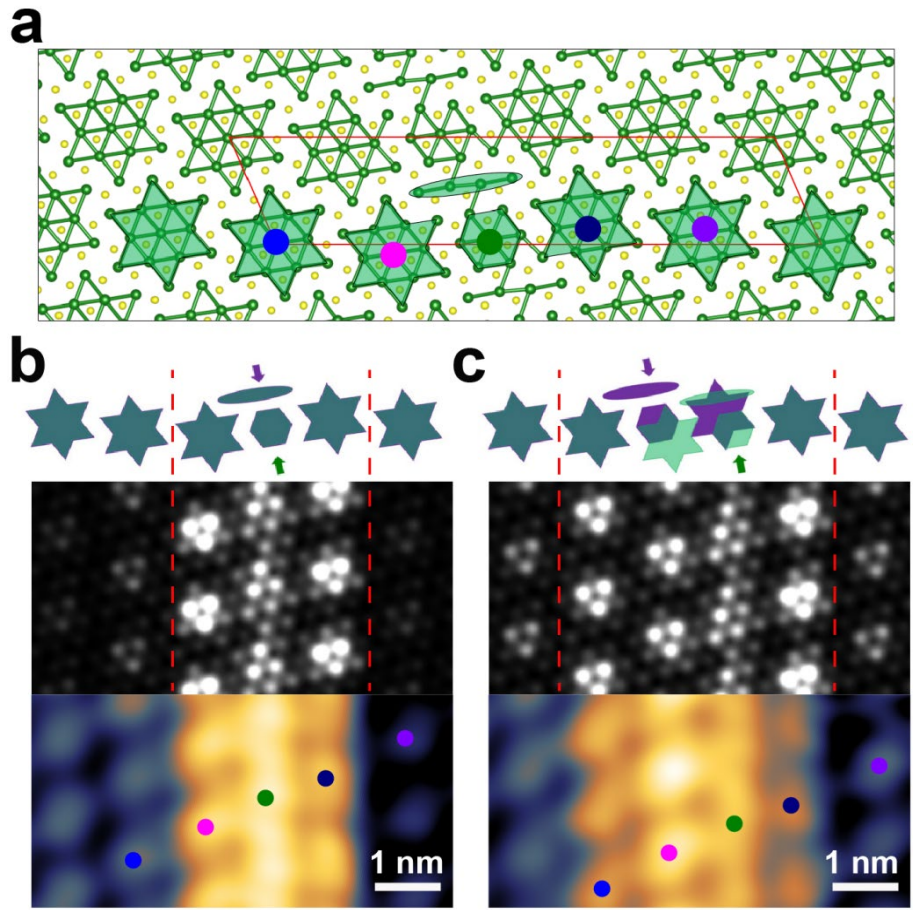

**Figure 2.** Structural model and simulated STM topography images. (a) Atomic structure of the top layer in bilayer DW-2 model. Each David-star cluster is marked by color dots. (b & c) Simulated STM images for no sliding and $a_{CCDW}$ sliding of the sublayer, respectively. Top panel is a schematic arrangement of the DS clusters (green: top layer and purple: sublayer) and the bottom panel is the experimental STM image ($V_b$ = 0.2 V).

### 2.3. LDOS of stacked DWs

Since the above topographic change comes from the emergence of the in-gap electronic states, the spectroscopy results can more directly prove it. The STS spectra and theoretical LDOS are shown for the clusters of ideally stacked and shifted DWs in **Figure 3**a and b, respectively. When the bottom layer DW is not shifted (Figure 3a), the CDW unitcells away from the DW and DB's (blue and purple dots) exhibit the well-known CDW gap. For these particular CDW clusters near the DW, the center of the gap shifts marginally by +0.08 eV due to the local band bending caused by the DW.[14,22] As mentioned above, the DW (green dot) and DB (pink and navy) clusters exhibit in-gap states above the Fermi energy at around +0.1 eV (arrows in the corresponding colors), which explains three bright rows observed in the empty-state STM image (Figure 2b). The theoretical LDOS (lower panel in Figure 3a) of the bilayer DW-2 model reproduces well these in-gap states (arrows in the corresponding colors. In the case of the bottom-layer DW shifted by $a_{CCDW}$ (Figure 3b) occur two prominent changes. Firstly, an in-gap state appears at the second nearest domain cluster to the DW (those with a blue dot). These clusters are hybridized electronically with those of the DW in the neighboring layer as mentioned above. Secondly, an additional in-gap state appears below the Fermi energy at around -0.05 eV at the DW (marked by black arrows in Figure 3b). This state is a resonance state with the in-gap state of the bottom layer DB (see supplementary Figure S3). Except for these two points, most of the other states in the top layer, including the in-gap states of DW and DBs, are similar to those of the pristine bilayer DWs'.

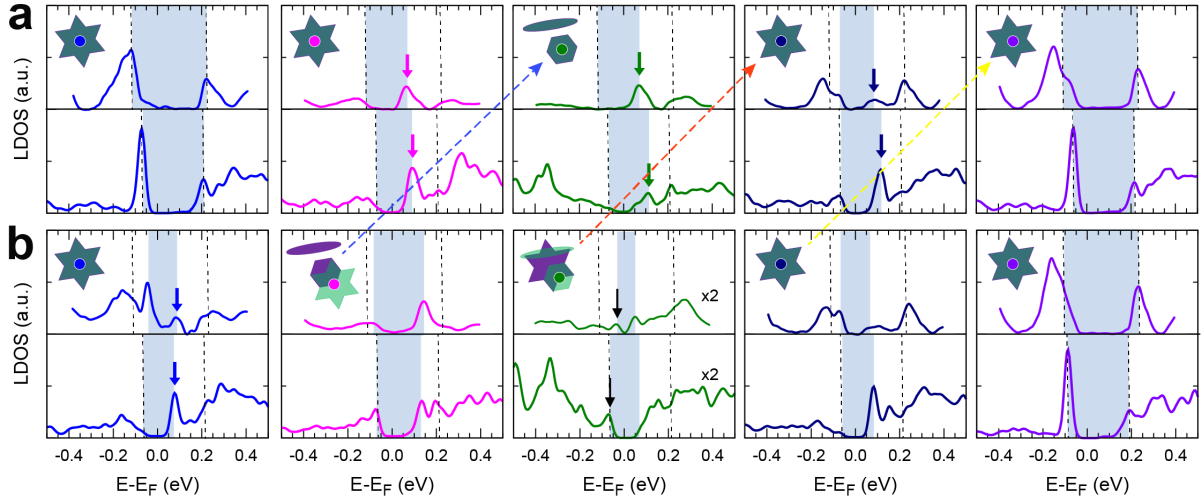

**Figure 3.** Local electronic density of states for two types of DWs. (a) Without sublayer DW sliding. (b) With sublayer DW sliding. The upper panel shows the experimental data while the lower panel is for theoretically calculated result. The pink, green, and navy STS curves are acquired from the LDB, DW, and RDB, respectively. The band gap of each cluster is represented by shades, while the domain regions are indicated by dashed lines, allowing for better comparison.

## 2.4. Kinkless 1D Electronic Junctions

The presence of diverse in-gap states along the DW and DBs leads to distinct local band gaps, which lead to distinct electronic junctions along the DW. For example, tunneling d$I$/d$V$ curves were measured at three different lines along the DW crossing the junction between DW-2A and DW-2B (dashed arrows in **Figure 4**a). These STS line scans (Figure 4b) reveal distinct electronic interfaces. Along the blue line in Figure 4a, one can find a junction of two semiconducting segments with slightly different band gap sizes and positions. The gap is changed in DW-2A since its in-gap state is shifted away from the Fermi energy (arrows on the left side in the right panel). On the other hand, the gap size is greatly reduced in DW-2A along the red line owing to the resonance state below the Fermi energy mentioned above (arrows on the left side in the middle panel). In the third case along the yellow line, one has a much bigger and abrupt discontinuity in the gap size; one moves from the DB clusters with an in-gap state in DW-2A (arrow in the right panel) to domain clusters without any in-gap state in DW-2B. The d$I$/d$V$ curves taken from spots #1-8 are demonstrated in the supplementary Figure S4. Note that the band gap change in the third case is abrupt at the unitcell level, while the changes in the other two cases involve interfacial states with a width of about two unitcells. In the first and second junctions, the unitcell structures of the top layer change from broken CDW clusters to perfect CDW clusters or vice versa. However, in the third case of an abrupt electronic junction, only the electronic structure changes without any accompanying structural deformation. While there are quantitative discrepancies on the band gap size, the above electronic interfaces are qualitatively explained by the DFT calculations with the double DW model of Figure 3. In Figure 4c, we plot further details of the band dispersions of the major in-gap states. They are consistent with the in-gap states of the line STS spectra (marked by black arrows). Note also that some of these segments (especially segments #1 and #6) feature strongly spin-polarized in-gap electronic states. These spin-polarized states suggest that the present junctions can be not only 1D electronic but also magnetic junctions.

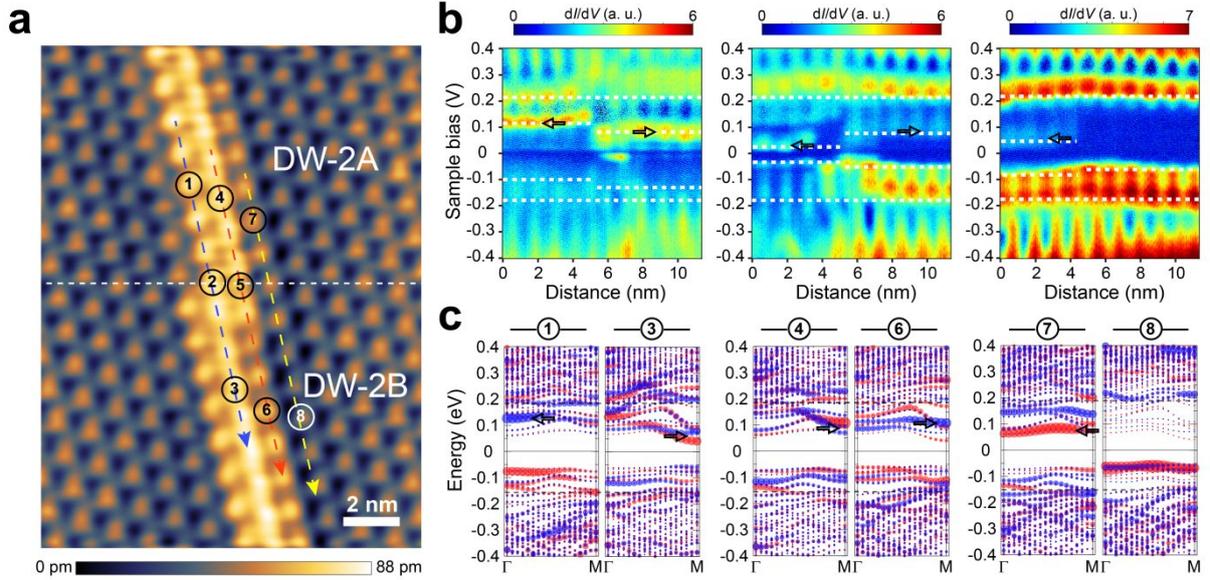

**Figure 4.** Spatial-resolved energy variation along the 1D DWs. (a) STM topography image including DW-2A and DW-2B, $V_b$ = 0.4 V. (b) Tunnelling conductance versus distance along the dashed arrows in (a), the left panel is for blue dashed arrow while middle panel and right panel are for red and yellow dashed ones, respectively. (c) Theoretically simulated energy band structure corresponding to (b).

## 3. Conclusion

We report on the purely electronic junctions identified along 1D domain-wall electronic channels on the 1T-TaS$_2$ surface via a combination of STM/STS measurements and DFT calculations. We demonstrate the existence of the stacked DW's in the top and second layers through the altered contrast in STM along the DW at a low scanning bias voltage. These DWs have different kink positions to produce local frustration of the stacking order of the CDW and DW unitcells. The change of the local stacking order induces in-gap states and local band gaps through the strong interlayer electronic coupling. We show that such local electronic variations through interlayer coupling produce a few distinct electronic junctions along the DW, one of which does not feature any intralayer structural interface. Our work indicates that the interlayer electronic coupling can effectively control the nanometer scale electronic property of 2D atomic monolayers.

## 4. Experimental Section

*STM and STS Measurements*: 1T-TaS$_2$ single crystal was grown by chemical vapor transport (CVT) technique. After cleaved in the load-lock chamber at room temperature under the pressure of 1×10-9 Torr, the 1T-TaS$_2$ sample was transferred into the scanning chamber for STM and STS measurements with a commercial STM (SPECS) in ultra-high vacuum at T = 4.4 K. The mechanically sharpened Pt-Ir wires were utilized as the STM tips. The differential conductance d$I$/d$V$ curves were recorded using a standard lock-in technique with a 10 mV bias modulation at frequency of 1 kHz.

*DFT Simulations*: We utilized the Vienna *ab initio* simulation package to perform density functional theory (DFT) calculations, implementing the Perdew-Burke-Ernzerhof functional and the projector augmented wave method.[25-27] We employed k-point mesh of 6×6×1 for Brillouin-zone integrations for the $\sqrt{13}\times\sqrt{13}$ Brillouin-zone integrations, and a plane-wave basis with a cutoff of 259 eV. The 1T-TaS$_2$ surface was modeled by a periodic slab consisting of two TaS$_2$ layers, with a vacuum spacing greater than 20 Å. We conducted atom relaxation until residual force components were within 0.02 eV, while maintaining the interlayer distance at the experimental value of 5.9 Å.[20] Additionally, we incorporated an on-site Coulomb energy of U = 2.3 eV for electron-correlation calculations.[28]

**Supporting Information**

Supporting Information is available from the Wiley Online Library or from the author.

**Acknowledgements**

Q. Y. and J.W.P. contributed equally to this work. This work was supported by the Institute for Basic Science.

Supporting Information

**Kinkless electronic junction along one dimensional electronic channel**

*Qirong Yao, Jae Whan Park, Choongjae Won, Sang-Wook Cheong, and Han Woong Yeom\**

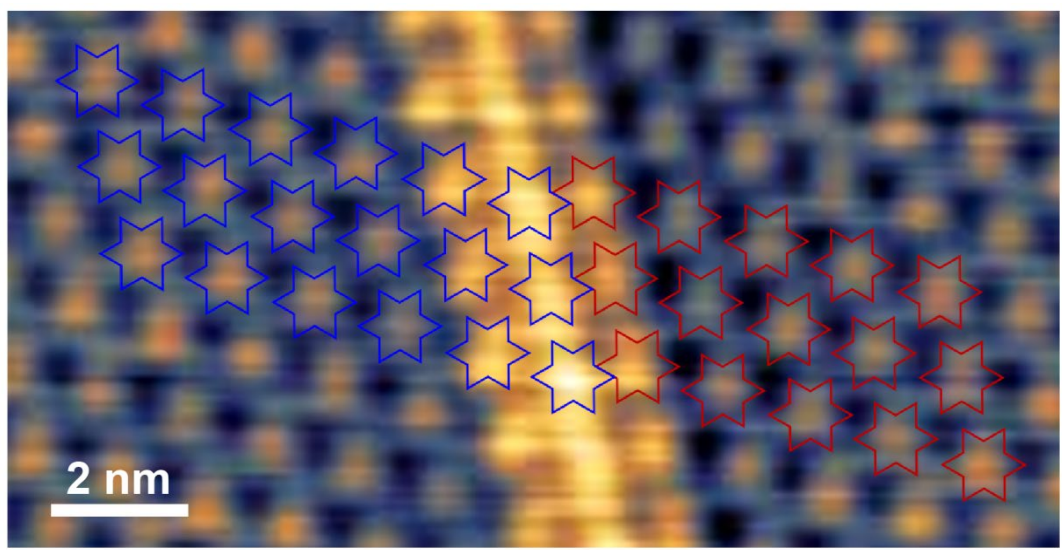

**Figure S1.** STM topography image with a second type of domain wall (DW-2).

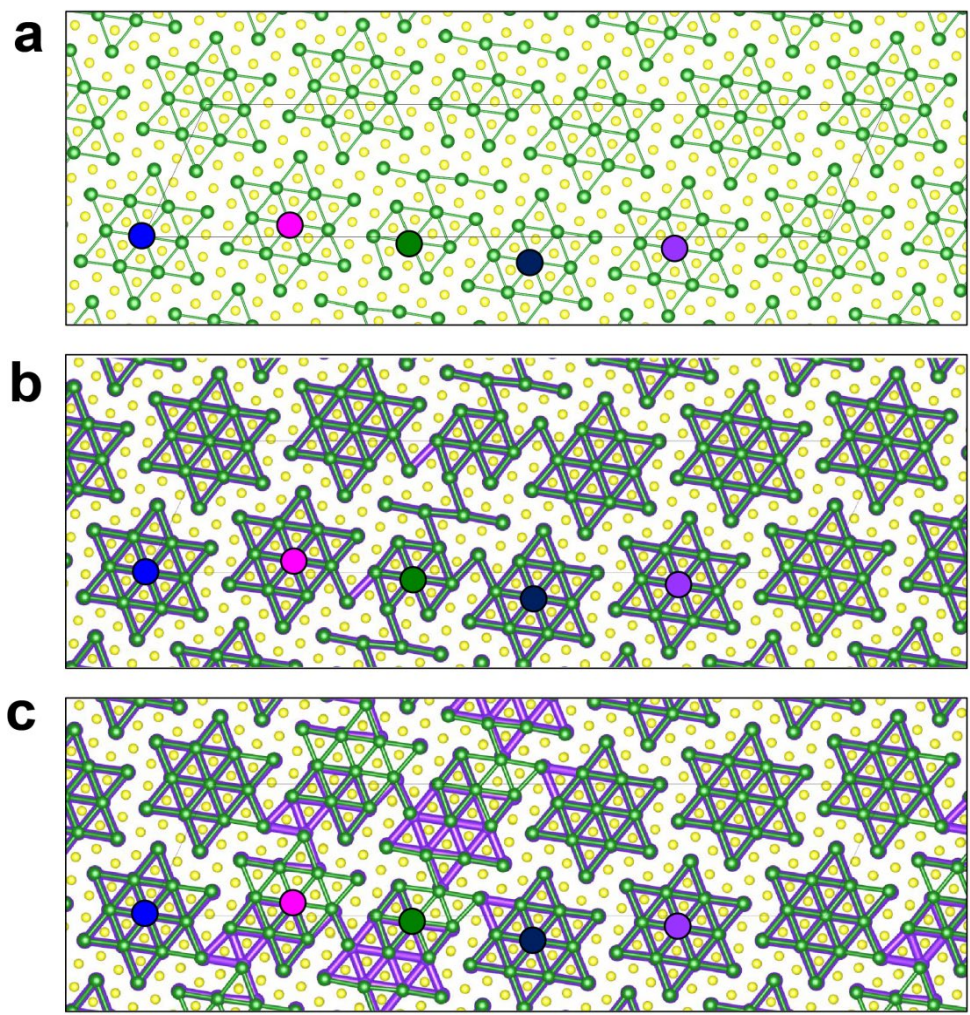

**Figure S2.** Atomic structure of the DW-2. (a) Single-layer model. (b & c) Bilayer model ((b) without; (c) with a$_{CCDW}$ sliding of sublayer DW). Green and purple balls represent the top and sublayer Ta atoms, respectively.

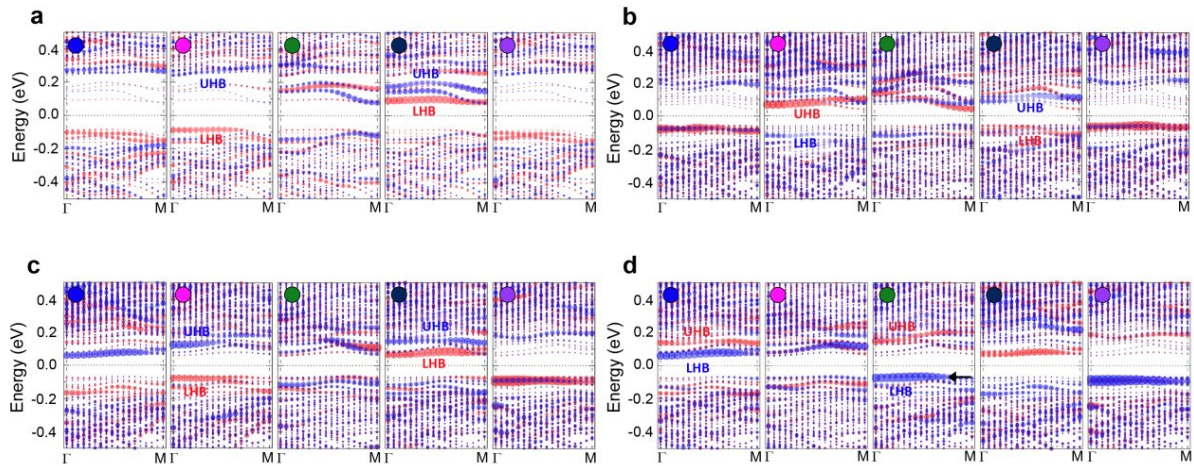

**Figure S3.** Band structures of the DW-2. (a) Single-layer model. (b-d) Bilayer model. ((b) without; (c & d) with $a_{CCDW}$ sliding of sublayer DW). (b & c) top layer, (d) sub layer. Red and blue circles denote the major and minority spins, respectively. Circle size is proportional to the states localized at corresponding CDW cluster in the Supplementary Figure S2. Arrow in (d) denotes the LHB state of the DS cluster at the sublayer below the DW. The energy level of the LHB corresponds to the energy of the additional in-gap state shown in Fig. 3b, indicating hybridized states with the sublayer.

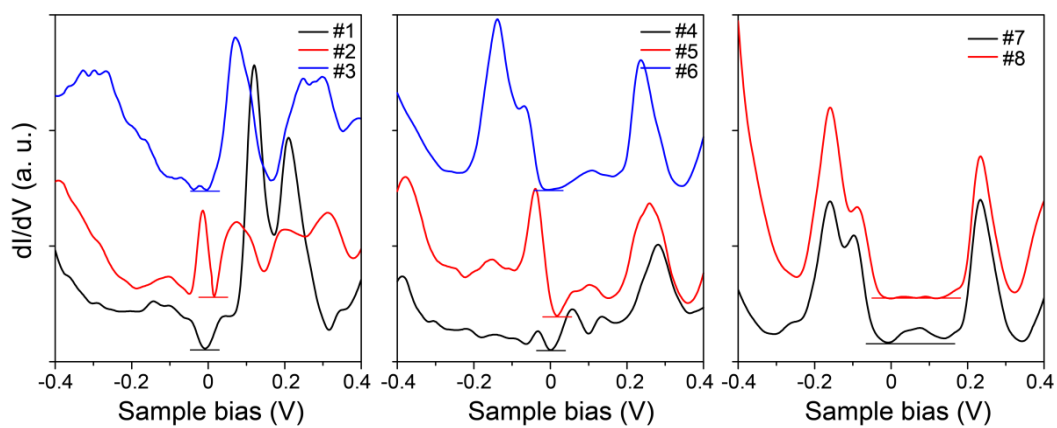

**Figure S4.** A series of d*I*/d*V* curves taken from the spots #1-8, which are marked in Figure 4a.